# From ultra-noisy to ultra-stable: optimization of the optoelectronic laser lock

Takuma Nakamura[1,2], Yifan Liu[1,2], Naijun Jin[3], Haotian Cheng[3], Charles McLemore[1], Nazanin Hoghooghi[1], Peter Rakich[3] and Franklyn Quinlan[1,4,5]

[1]Time and Frequency Division, National Institute of Standards and Technology, 325 Broadway, Boulder, CO 80305, USA
[2]Department of Physics, University of Colorado Boulder, 440 UCB Boulder, CO, 80309, USA
[3]Department of Applied Physics, Yale University, New Haven, Connecticut 06520, USA
[4]Electrical, Computer and Energy Engineering, University of Colorado, Boulder, Colorado 80309, USA
[5]fquinlan@nist.gov
*takuma.nakamura@nist.gov

**Abstract:**

We demonstrate thermal-noise-limited direct locking of a semiconductor distributed feedback (DFB) laser to a sub-1 mL volume, ultrastable optical cavity, enabling extremely compact and simple ultrastable laser systems. Using the optoelectronic laser locking method, we realize over 140 dB suppression of the DFB free-running laser noise at 10 Hz offset, a level we estimate to be ~ 70 dB greater than Pound-Drever-Hall locking can provide, and reach a phase noise level of -120 dBc/Hz at 200 kHz offset. We also demonstrate a new feedforward noise correction method that improves the quality of the heterodyne beat with an optical frequency comb by providing another 60 dB of laser noise rejection – a level that is 15 dB greater than predicted by current models. With feedforward, we transfer the cavity thermal noise limit across the comb spectrum despite the fact that the cavity-locked laser itself is noisy. These results establish a simple, low noise, compact approach to ultrastable laser locking that is compatible with integrated photonics, with applications in low phase noise microwave generation, sensing, and satellite ranging.

## 1. Introduction

Ultra-stable optical frequency references are vital components in optical atomic clocks [1], low noise microwave generation via optical frequency division [2–4], gravitational wave detection [5], environmental sensing [6,7], and quantum networking [8,9], with applications ranging from geodesy [10], global hydrodynamics [11], and earthquake detection [12], to microwave spectroscopy [13] and synthetic aperture radar [14,15]. To achieve the requisite level of performance, lasers need to be stabilized to a frequency reference, such as a Fabry-Pérot (FP) cavity, fiber delay line [16], or integrated on-chip resonator [17,18]. Although many methods have been demonstrated to stabilize lasers to references [19–24], Pound-Drever-Hall laser locking (PDH) has been the gold standard for more than 40 years. This is due to its numerous advantages, including robustness and large signal-to-noise ratio, that allow for laser stabilization to less than $10^{-5}$ of the width of a resonance of a reference cavity, resulting in millihertz linewidth lasers [25].

Despite its excellent track record, the PDH locking method has disadvantages, particularly for compact ultrastable laser systems. Successful locking to a high-finesse cavity typically requires a laser linewidth that is narrow to begin with. For example, kHz-level linewidths from fiber lasers, extended cavity or external-feedback diode lasers, nonplanar ring oscillator (NPRO) lasers, or otherwise pre-stabilized lasers are required to reach Hz-level laser stability [26]. With out-of-the-lab applications of ultrastable lasers moving toward compact chip-scale solutions, direct PDH locking becomes more challenging, as chip-scale lasers tend to have much larger linewidths. Furthermore, limited feedback gain at high offset frequencies results in unsuppressed residual laser noise, forming a "servo bump" in excess of the cavity noise limit. Such noise bumps are detrimental to optically derived low noise microwave signals.

As an alternative method to overcome these drawbacks, optoelectronic oscillator (OEO) laser locking offers a compelling approach. First proposed to the best of our knowledge in 1998 and 2000 [27,28], this method was revived in 2024 with new measurements, the addition of an optical feedforward noise correction and a thorough theoretical analysis [29]. In the OEO laser locking method, the laser phase noise is written onto the phase of an RF (radio frequency) carrier that oscillates in a regenerative OEO loop. The transference of the laser phase noise to the RF domain has the advantages of 1) enabling a feedback-locking servo that is proportional to phase fluctuations to provide greater noise suppression, and 2) facilitating simple-to-implement feedforward noise correction, where the phase noise of the laser is mixed with the phase noise on the OEO-generated RF carrier. In [29], this feedforward was implemented by correcting the laser phase with an AOM that was driven by the OEO carrier.

In this paper, we demonstrate several new capabilities and extreme noise suppression with an OEO laser lock to a sub-1 mL volume FP resonator that make significant advances towards integrated photonics compatibility. First, we demonstrate a new feedforward correction scheme amenable to a large class of applications that rely on heterodyne detection with another optical source, such as optical frequency comb stabilization, microwave generation, or laser frequency-based sensing. With this new and simple RF feedforward noise correction scheme, we demonstrate locking of an optical frequency comb to an OEO-locked laser with high residual noise, where the resultant comb noise is nonetheless at the FP cavity thermal noise limit. We further show that careful tuning of the OEO oscillation frequency results in feedforward-corrected laser noise suppression that is 15 dB larger than current theory predicts. Lastly, we show OEO laser locking can stabilize lasers with much higher noise than a PDH lock can handle, where the free-running linewidth of the laser is ~550 kHz. Feedback control of this laser produces >140 dB of phase noise suppression at 10 Hz offset. We estimate this is ~70 dB greater noise suppression than is possible with PDH using the same servo electronics, though robust PDH locking was in fact not possible with this laser. Feedforward correction then further reduced the effective noise ~35 dB at 1 MHz and >20 dB at 5 MHz offset from the carrier frequency. By combining a compact laser source with our compact reference cavity, these results demonstrate a simple, low noise, cost effective approach to ultrastable laser locking that is compatible with integrated photonics.

## 2. Principle of Operation

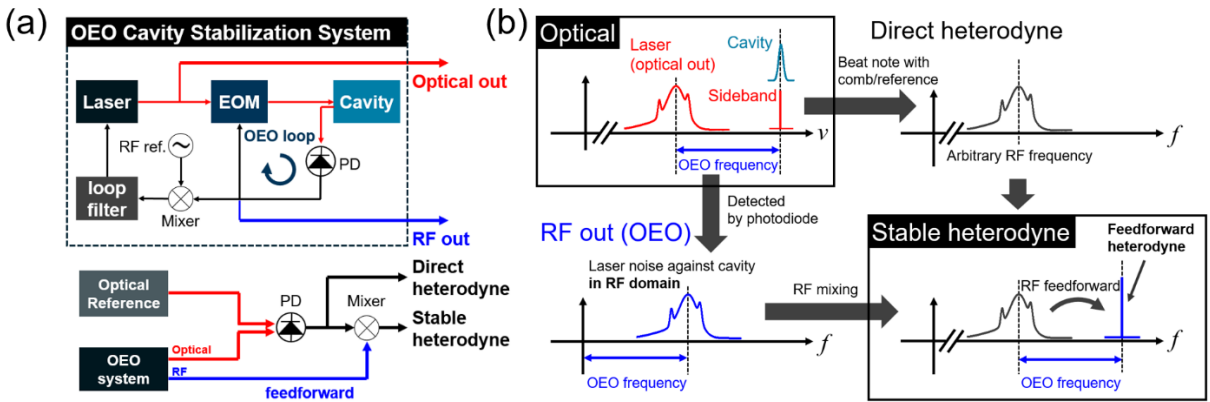

Fig. 1 (a) Schematic of the OEO laser lock (top) and RF feedforward correction (bottom). (b) Frequency domain representations of the laser, cavity mode and resonant sideband (top left), OEO frequency (bottom left), direct heterodyne beat with an optical reference (top right) and feedforward-corrected heterodyne (bottom right).

In this section, we briefly review the operating principle behind the OEO laser lock, and we describe how its operation allows for low noise control of optical frequency combs or other heterodyne beat signals. A schematic of the locking setup is shown in Fig. 1. Just as with a PDH lock, a continuous-wave (CW) laser is first modulated by an electro-optic phase modulator (EOM), generating discrete modulation sidebands. Unlike PDH, the modulation is created by a regenerative optoelectronic (OEO) feedback loop, starting from noise. After passage through the EOM, the laser illuminates an ultrastable optical frequency reference with the laser frequency detuned from the nearest resonant optical mode. The reflection from the cavity is then directed to a photodetector. The signal from the photodetector is amplified and fed into the RF modulation port of the EOM, forming the regenerative OEO loop. Note that the frequency detuning between the laser and the cavity mode should be within the photodetector and EOM modulation bandwidth.

At the beginning of OEO oscillation, the laser is modulated with white noise, creating broadband phase modulation of the laser. The frequency component of this modulation that is resonant with the optical cavity experiences a phase shift upon reflection, converting the original phase modulation to amplitude modulation that can be photodetected. If the detuning between the laser frequency and the cavity mode is also a resonance of the OEO loop, and the loop gain is sufficient, self-sustaining OEO oscillation occurs, creating a discrete sideband that is aligned with the mode of the optical cavity.

The OEO loop keeps the OEO modulation sideband on resonance with the optical reference cavity despite fluctuations in the optical carrier frequency. When the laser frequency shifts, if the modulation frequency is fixed, the sideband frequencies will shift commensurately. However, light stored in the optical cavity leaks out and heterodynes with the laser carrier. This shifts the frequency of the beat that defines the OEO oscillation frequency, thereby shifting the EOM modulation frequency. The adaptive shift in the EOM modulation frequency compensates fluctuations in the laser frequency to keep the sideband on resonance with the optical cavity. The shorter the OEO loop delay, the faster the phase modulation frequency will shift to compensate. Similarly, the longer the cavity storage time, the stronger the restoring force to keep the sideband on resonance. Since the sideband stays resonant with the optical cavity, the laser phase noise (relative to the optical cavity mode) is transferred to the OEO oscillation frequency.

The transference of the laser phase noise to the OEO signal can be exploited for feedback and correction of the phase noise of the laser [28], yielding a large reduction in the laser noise. Feedback to the laser also prevents large deviations on the laser frequency that would quench OEO oscillation. As discussed in more detail in the Supplement, in the high gain limit, the noise reduction of this phase lock is greater than that of PDH due to the fact that, for PDH, the error signal is proportional to frequency rather than phase. As we show below, this allows for robust locking of lasers that have much higher free-running phase noise while still reaching the cavity thermal noise limit.

The fidelity by which the laser phase noise power spectral density is transferred to the OEO frequency depends on the cavity storage time and the propagation delay within the OEO loop, and can be calculated through a linearized model (see Refs. [28,29] and Supplement). The ratio of the phase noise power spectral density (PSD) of the OEO frequency, $S_\varphi^{OEO}(f)$, to the laser phase noise PSD, $S_\varphi^L(f)$, is given by

$$S_\varphi^{OEO}(f)/S_\varphi^L(f) \approx \left(\frac{\tau_{cav}}{\tau_{cav}+\tau_{OEO}}\right)^2 \quad (1)$$

Here, $\tau_{OEO}$ is the propagation time around the OEO loop excluding the optical reference cavity, and $\tau_{cav}$ can be interpreted as the group delay of the cavity or the storage time of the optical field (twice the cavity intensity ringdown time). As $\tau_{cav}$ can exceed $\tau_{OEO}$ by several orders of magnitude, the OEO frequency gives an excellent approximation of the laser phase noise.

The residual laser phase noise can be further suppressed in a simple manner with feedforward correction, particularly important at large offset frequencies where the feedback gain is reduced [29]. Feedforward correction subtracts the phase noise of the OEO loop frequency from the laser, such that the level of noise suppression with feedforward is given by

$$(S_\varphi^L - S_\varphi^{OEO})/S_\varphi^L = \left(1 - \frac{\tau_{cav}}{\tau_{cav}+\tau_{OEO}}\right)^2 = \left(\frac{\tau_{OEO}}{\tau_{cav}+\tau_{OEO}}\right)^2 \approx \left(\frac{\tau_{OEO}}{\tau_{cav}}\right)^2 \quad (2)$$

With $\tau_{cav}$ of a high finesse cavity as high as tens of μs and with a $\tau_{OEO}$ of tens of ns, Eq. (2) implies feedforward correction can be greater than 60 dB for large, laboratory-based cavity systems. For compact, 1 mL-volume cavities, feedforward correction can still be greater than 40 dB. Feedforward correction can be implemented by actuating on the laser with an AOM, as demonstrated in [29]. However, for many important applications of cavity-stabilized lasers, such as frequency stability transfer with an optical frequency comb, feedforward does not need to be applied in the optical domain. Rather, feedforward correction can be applied to the heterodyne beat between the OEO-locked laser and the frequency comb or other laser source. Implementing this RF feedforward is performed by simply frequency-mixing the OEO oscillation frequency with the heterodyne beat. Such RF feedforward has the advantages that no additional electro-optic components are required, the OEO frequency need not match the resonant frequency of an AOM, and the correction bandwidth can be MHz wide. This is particularly advantageous when locking lasers with higher broadband noise, such as standard semiconductor lasers. It is important to note that feedforward correction is possible with PDH locking as well by taking the voltage error signal, finely tuning the voltage gain, and applying it to an electro-optic actuator [30,31]. However, feedforward with the OEO lock has the distinct benefits of a correction signal that is encoded in the phase of an RF carrier rather than a baseband voltage (such that no gain control is necessary), simple implementation with a frequency mixer, and, as shown in detail below, large noise suppression over a large bandwidth.

Lastly, it is important to note that the linearized model fails to capture high-power effects, such as higher-order EOM phase modulation sidebands and the generation of harmonics of the OEO frequency, that may alter the relationship between the laser and OEO frequencies. Below we show that, for a system appropriately tuned, the feedforward correction can be much greater than this model predicts.

## 3. Experiment and Results

We conducted four series of experiments on the OEO laser lock with the RF feedforward scheme and its noise suppression capacity: a comparison between PDH and OEO locking of a narrow linewidth fiber laser; optical frequency comb stabilization with RF feedforward; nonlinearity and feedforward noise suppression investigations; and OEO locking of a much larger linewidth distributed feedback (DFB) laser. For all experiments, the lasers were locked to an in-vacuum-bonded Fabry-Pérot "minicube" cavity made of ultralow expansion (ULE) glass. The cavity is manufactured with a micro-fabricated mirror and is diced out of a cavity array. In-vacuum bonding of the cavity mirrors and spacer provides a vacuum gap between the mirrors while the cavity is operated in air, resulting in thermal noise limited performance in a compact package. The cavity spacer length is 4 mm with a corresponding free-spectral range of 37.5 GHz, the finesse is about 540,000, and the full width of the resonant modes near 1550 nm is ~70 kHz such that the resonator quality factor is ~2.7 billion. The cavity volume is less than 1 mL, and it is placed in an air-tight enclosure measuring 50 mm x 50 mm x 30 mm. More details of the cavity construction and performance may be found in [32,33].

*3.1 Comparisons between OEO and PDH locking*

First, we compared the phase noise performance of PDH and OEO laser locking using a narrow linewidth 1550 nm fiber laser. Locking with PDH followed the standard scheme [34], where we utilized a 29 MHz source to sinusoidally modulate the laser phase, generating a voltage error signal that is proportional to the difference between the laser frequency and the frequency of the FP cavity resonant mode. The error signal voltage is then conditioned by a loop filter servo with proportional, integral and differential gain, the output of which is used to correct the laser frequency. Fast frequency corrections were performed via an AOM external to the laser while slow, large dynamic range corrections were implemented using a piezoelectric transducer (PZT) to change the laser cavity length. For the OEO lock, the phase modulation source was replaced by the regenerative loop, as shown in Fig. 2(a). The OEO loop delay is 25 ns, determined by measuring the frequency separation between the fundamental OEO resonant frequency and its harmonics. With a loop delay of 25 ns and a FP cavity half-linewidth of 35 kHz, from Eq.(2) the anticipated laser noise rejection with RF feedforward is ~45 dB. Importantly, no electrical filters are included in the OEO loop, as their inclusion can significantly increase the loop delay. Any mode within the photodetector bandwidth may be chosen as the OEO oscillation frequency by appropriately detuning the laser frequency from the FP cavity mode. For the experiments reported in this section, the OEO frequency was chosen to be 85 MHz. Feedback correction to the laser was implemented by frequency-mixing the OEO frequency with a reference oscillator. The error signal then corrects the laser fluctuations using the same loop filter servo settings and feedback actuators as the PDH setup. Additionally, for fairest comparison, the optical power was kept the same between the two setups, as was the power of the RF signal applied to the EOM. In contrast to PDH, the resulting voltage error signal is proportional to the phase error of the laser.

Resulting phase noise measurements are shown in Fig. 2(b). Out-of-loop phase noise measurements were performed by heterodyning against independent ultrastable lasers and utilizing the cross-correlation measurement technique [35]. This method reveals the noise of the laser under test by rejecting the noise of the reference lasers and was particularly important for noise offset frequency greater than ~1 kHz. For PDH, the in-loop, or residual, laser phase noise is determined by measuring the voltage noise of the error signal and calibrating against the slope of the discriminant [36]. For the OEO laser lock, the in-loop noise is the phase noise of the OEO carrier frequency when the feedback to the laser is engaged.

The free-running phase noise of the fiber laser is shown in the dotted grey curve in Fig. 2(b). To determine the free-running fiber laser linewidth, we applied the beta separation-line method [37]. Integration of the noise to 1 Hz yielded a linewidth of 12 kHz. When locked using PDH, the phase noise of the laser output is reduced, yielding the dark red curve, and follows the cavity noise limit for offset frequencies below ~10 kHz. For larger offset frequencies, the out-of-loop noise increases above the cavity noise limit, forming a "servo bump". Comparison to the residual noise curve (light red) confirms the servo bump is a result of unsuppressed free-running laser noise due to the reduced gain in the feedback control circuit at higher offset frequencies. Note that the PDH residual noise shown here is characterized by the frequency discriminator slope with a correction at higher frequency offset, since the PDH error acts as phase discriminator for offset frequencies outside the cavity linewidth. More information can be found in ref [38] and the Supplement.

Exchanging the PDH circuit with the OEO laser lock with RF feedforward results in the out-of-loop phase noise shown in Fig. 2(b) in dark blue. As with PDH, the phase noise follows the cavity limit for low offset frequencies. Without feedforward correction, the noise at offset frequencies above 10 kHz would show a similar servo bump as PDH, because the feedback bandwidth is limited in the same way as PDH, as indicated by the residual OEO loop noise (light blue curve). RF feedforward correction on the heterodyne beat eliminates the servo bump, reducing the noise at 100 kHz by ~20 dB to a floor near -110 dBc/Hz. Techniques that further suppress this floor are discussed in Section 3.3. It is also interesting to compare the residual laser noise of PDH and OEO locking at low offset frequencies, where, even without feedforward correction, the OEO lock provides 120 dB of laser noise suppression at 10 Hz offset, 50 dB more than PDH. This implies source lasers with much higher free-running noise can be locked with the OEO method, as shown in Section 3.4.

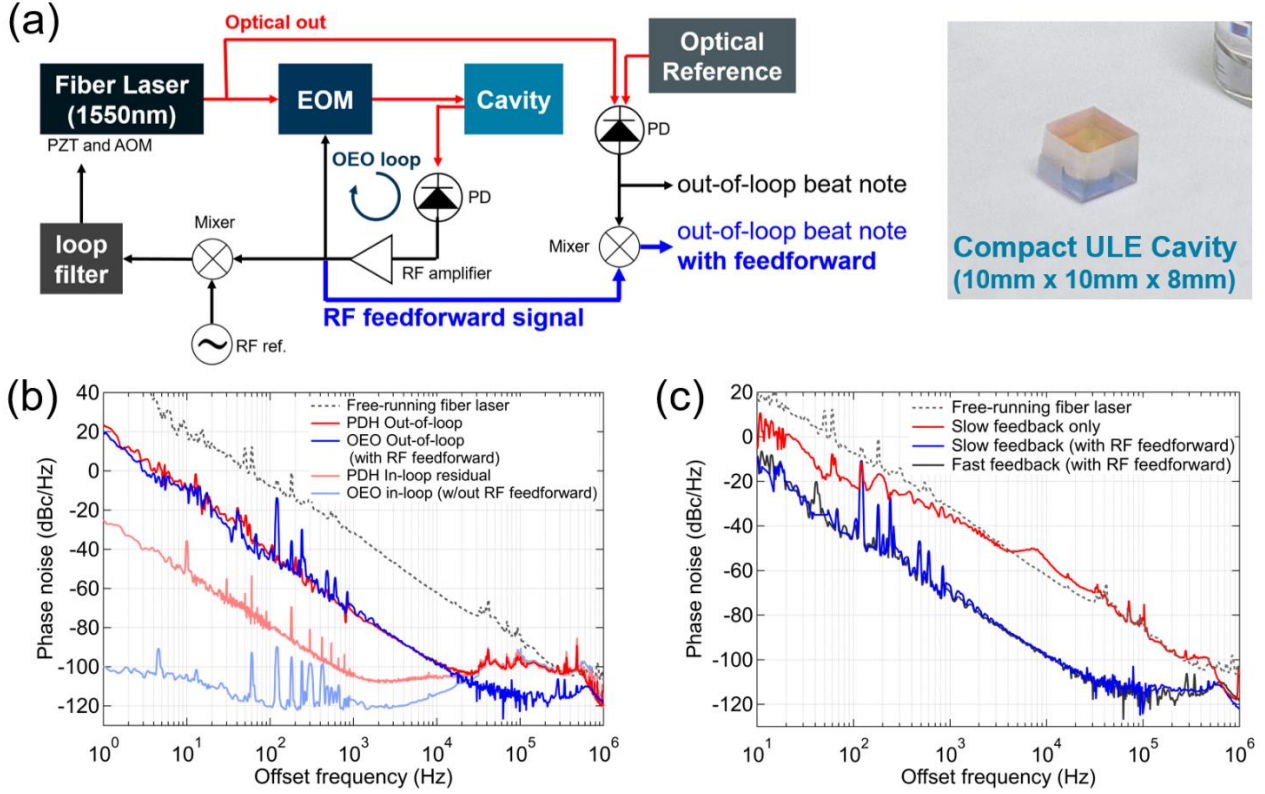

Fig. 2. (a) OEO laser lock schematic with fiber laser and minicube reference cavity. (b) Phase noise spectra with high bandwidth feedback using an AOM, comparing PDH and OEO locking with RF feedforward. Both locking methods use the same laser, loop filter, AOM for feedback control and optical reference cavity. The OEO lock with RF feedforward follows the cavity noise limit for offset frequencies below ~ 100 kHz. The fiber laser's free-running noise, and the in-loop residual noise of the laser for both PDH and OEO lock are shown as well. (c) Comparison of the phase noise of the fiber laser with slow, PZT-only feedback control with and without RF feedforward. With RF feedforward, the phase noise is indistinguishable from the phase noise where high bandwidth AOM feedback is used.

To further demonstrate the ability of feedforward correction, we increased the residual fiber laser noise by removing the AOM from the feedback circuit, leaving only the low-bandwidth PZT actuator. This greatly decreased the feedback bandwidth and increased the laser noise, as shown in the out-of-loop beat note in Fig. 2(c). Correction to the beat note with RF feedforward reduced the noise by more than 45 dB, restoring the phase noise to the cavity limit. This level of noise rejection is larger than expected from Eq. (2). Further experiments helped clarify this discrepancy and are described in Section 3.3.

*3.2 Stability transfer to an optical frequency comb via RF feedforward*

With the OEO lock, RF feedforward correction allows for the transference of thermal noise-limited cavity stability to an optical frequency comb (OFC) despite the fact that the optical signal itself can have excess laser noise. Locking an OFC to a cavity-stabilized laser typically requires the formation of a heterodyne beat between the cavity-stabilized laser and a comb tooth of the OFC. This beat note is then used in a phase-lock loop to stabilize the comb to the cavity-stabilized laser. To demonstrate the utility of RF feedforward, we phase locked a home-built octave-spanning Er:fiber-based optical frequency comb to the OEO-locked fiber laser, the schematic of which is shown in Fig. 3(a). The OFC's offset frequency is stabilized with a conventional *f-2f* interferometer, and the comb contains an intracavity EOM for tight phase locking to optical references [4]. The fiber laser was locked to the minicube cavity with PZT feedback only, such that the laser noise again was poorly suppressed by the feedback loop. The optical frequency comb was then tightly locked to the fiber laser output via the internal EOM with and without RF feedforward. We then assessed the transfer of the cavity stability to the comb by measuring the phase noise of a beat note between another comb tooth at 1156 nm and a separate ultrastable laser that is locked to a 30 cm long FP cavity with PDH [39].

Results are shown in Fig. 3(b). Without RF feedforward, the high residual noise of the fiber laser is transferred to the comb as expected, as shown in the red trace. On the other hand, when the RF feedforward was implemented, the phase noise of the comb at 1156 nm is greatly reduced, reaching the cavity thermal noise limit for offset frequencies below 2 kHz. Above 2 kHz offset, the measurement is limited by noise on the 1156 nm reference laser. In addition to low noise transfer across the optical spectrum, such locking of an OFC can enable low noise microwave generation via optical frequency division while reducing the noise requirements of the free-running reference laser [2].

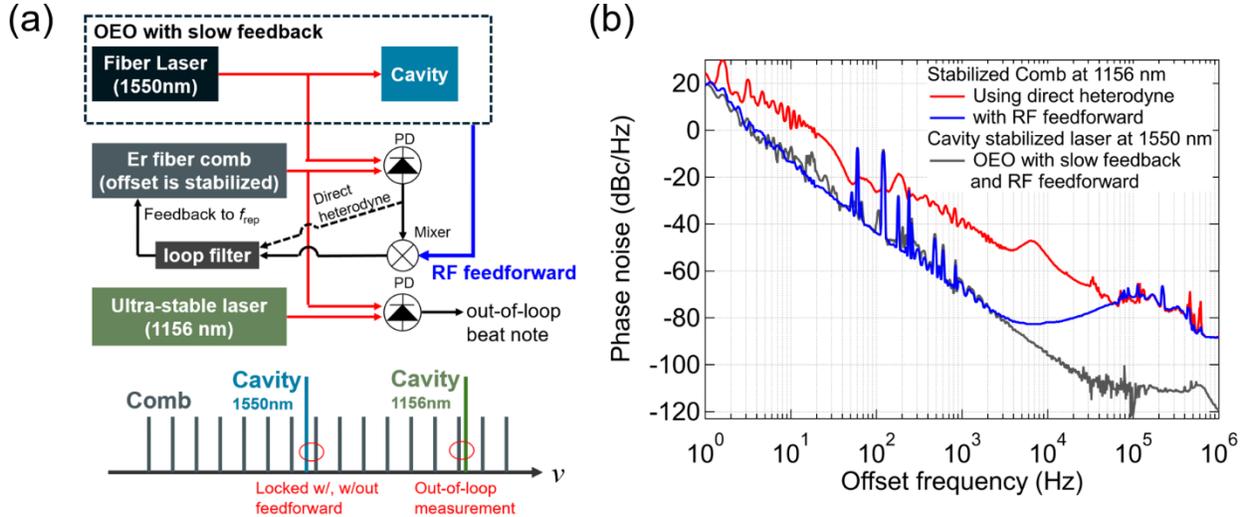

Fig. 3. Optical frequency comb locking to a cavity using the OEO laser lock. (a) Schematic (top) and frequency domain view of the optical frequency comb lock. (b) Phase noise spectra of the beat note between the frequency comb and the 1156 nm optical reference with and without the use of RF feedforward on the 1550 nm beat note between the comb and the OEO-locked laser. The phase noise of the 1550 nm laser from Fig. 2(c) is scaled to 1156nm for comparison.

*3.3 OEO laser lock nonlinearities and enhanced laser noise suppression.*

While low phase noise is realized over a broad range of system parameters, we find enhanced performance can be achieved by fine-tuning the OEO lock operation. We first show distinct operation regimes as a function of the input optical power, where each regime exhibits a different OEO amplitude and phase noise level at large offset frequencies. By operating in a specific nonlinear regime, we lower the phase noise of the feedforward-corrected signal for offset frequencies >10 kHz to ~ -120 dBc/Hz. Second, we show that feedforward correction is able to suppress the laser noise beyond what is given by Eq. (2) by properly tuning OEO oscillation frequency.

We varied the laser power to the cavity with a variable optical attenuator and have identified three distinct oscillating regimes, each with unique noise properties. Results are summarized in Fig. 4. Figure 4(a) shows the OEO power driving the EOM as a function of the optical power. Once the optical power reaches the threshold needed for OEO oscillation, the OEO power increases as the optical input power is increased, denoted as Region 1 in the Figure. Note that the optical powers described here to reach OEO oscillation and the reported OEO power can vary depending on the total gain of the amplifiers in the loop and V-pi of the EOM. In our case, the gain of the transimpedance amplifier following the photodiode is 10 kV/A, the RF amplifier has a gain of ~20 dB, and the EOM has a V-pi of ~3.5V. At an optical power of 0.6 mW, the OEO power reaches a maximum of 11.4 dBm, after which it decreases with increasing optical power (Region 2). At 1.4 mW of optical power, the OEO power again begins to increase (Region 3). Transition from one region to the next is accompanied by a sudden change to the noise spectrum of the OEO frequency, with representative spectra shown in Fig. 4(b). Notably, the noise of Region 1 is highest, and has asymmetric noise sidebands that imply correlation between amplitude and phase noise of the OEO carrier [40]. Representative amplitude and phase noise measurements of the OEO carrier for each region of operation are shown in Fig. 4(c). The relative intensity noise (RIN) of the laser is also included in Fig. 4(c) for comparison. A noise bump near 600 kHz is common to all amplitude noise plots, the frequency of which matches the relaxation oscillation noise bump of the laser RIN. This noise peak is also present in the phase noise, implying amplitude-to-phase noise coupling, with Region 2 providing the lowest noise. Thus, while feedforward correction relies on the OEO loop to generate a faithful copy of the laser phase noise, we believe that amplitude noise, especially at higher frequency offset, is converted to phase noise of the OEO signal. This excess OEO phase noise is not on the direct laser output, and it will not be canceled with feedforward correction. This phenomenon currently restricts the achievable phase noise limit with OEO lock. By operating in Region 2, we reduce both the amplitude and the phase noise of the OEO signal, leading to lower phase noise after RF feedforward, shown in Fig. 4(d). Noise from 100 kHz to 1 MHz is reduced 5-10 dB to ~ -120 dBc/Hz. This represents the lowest phase noise in this offset frequency range of a PDH- or OEO-locked system of which we are aware.

The linearized noise model for the OEO lock gives a simple prediction of the fidelity by which the laser noise is transferred to the OEO frequency, and thereby the amount of laser noise rejection feedforward can provide. As stated above, for our cavity linewidth and OEO loop delay, we estimate the feedforward noise suppression ratio to be ~45 dB. To verify this experimentally, we employed two separate measurement methods. The first method consisted of a 5 MHz-wide sweep of the reference synthesizer to which the OEO frequency was phase locked. Although self-sustaining OEO oscillation is centered at specific frequencies, the short OEO loop delay results in broad resonances over which the OEO frequency may be tuned [29]. Since the OEO phase lock is enabled by actuating on the laser frequency, sweeping the OEO phase reference results in a laser frequency sweep of nearly equal amount. Therefore, by comparing the change of the feedforward-corrected heterodyne beat to the change in the OEO frequency, the level of laser noise suppression can be estimated. As shown conceptually in Fig. 5(a), if the OEO frequency is changed by 1 MHz, the laser frequency should also change by 1 MHz. With feedforward, however, the laser frequency shift is largely canceled, resulting in a much smaller kHz-level shift. The ratio of the shift on the feedforward-corrected signal to that of the laser corresponds to the noise suppression ratio. In Fig. 5(b), the 5 MHz OEO frequency detuning is suppressed to about 35 kHz with feedforward correction. The noise suppression ratio, given by the average slope of the curve in Fig. 5(b), corresponds to around -46 dB, consistent with the theoretical prediction. However, there is a region of near-zero slope around 0 MHz detuning, implying that the laser noise should be perfectly cancelled by feedforward. To verify this, a second noise suppression measurement method was implemented. The bandwidth of the laser phase lock was reduced to less than 10 kHz, and the laser was externally frequency modulated at 20 kHz with an AOM. The phase noise power of this tone was measured with and without RF feedforward, as shown in Fig. 5(c). We measured the amount of suppression of this tone as a function of OEO frequency tuning and compared it to the slope of the frequency sweep measurement. As shown in Fig. 5(d), the agreement between these measurement techniques is excellent, not only matching the overall noise suppression level but also some of the fine structure. The noise suppression in this measurement is limited to around 60 dB, which is mainly limited by the delay error of the feedforward correction as discussed in Section 3.4 and the Supplement. While larger feedforward suppression may be achieved with fine tuning the feedforward delay, the 20 kHz tone is already suppressed to a level very close to the measurement floor (determined by the cavity noise), such that demonstrating >60 dB suppression is difficult with the current setup. Efforts to fully understand the existence of noise suppression beyond that predicted in Eq. (2) are underway. However, we find that the power in the second harmonic of the OEO oscillation frequency is correlated with regions of high noise suppression, indicating the importance of higher-order terms not included in the noise model. This is discussed further in the Supplement.

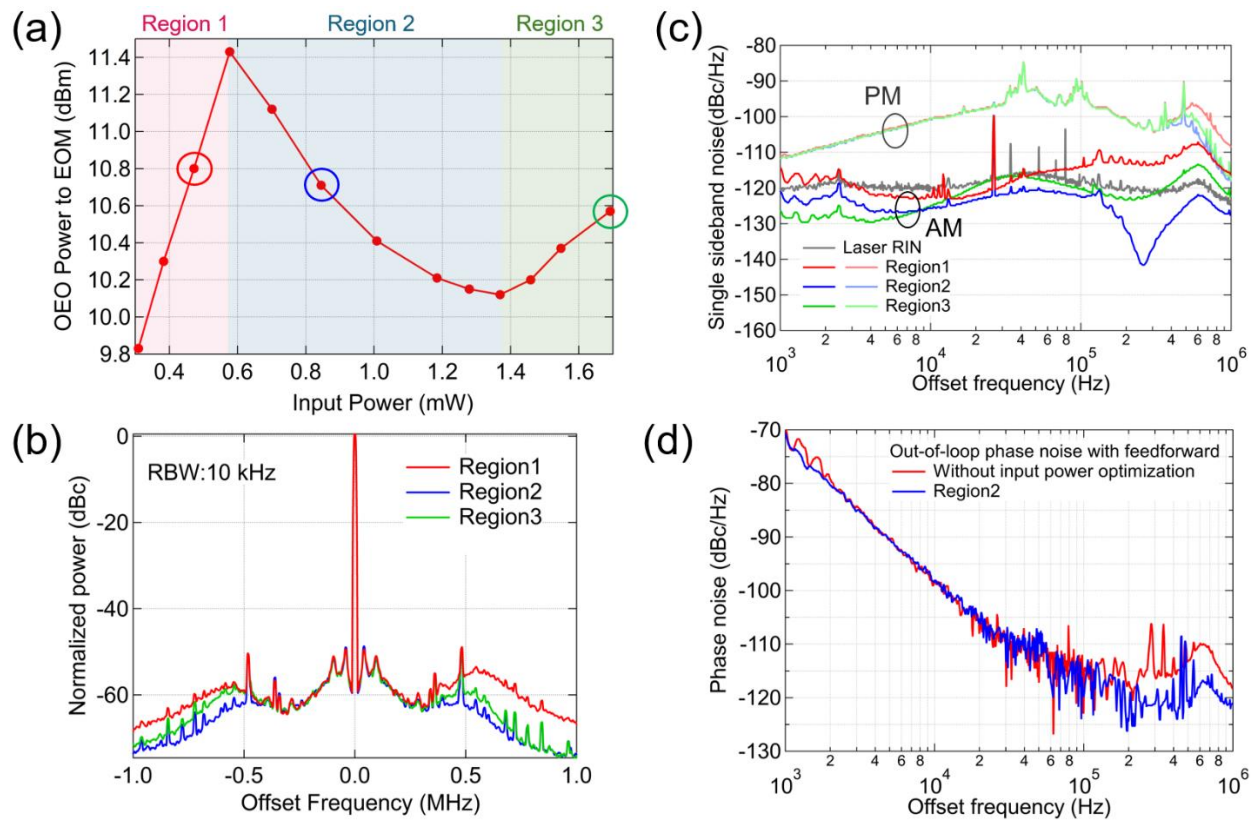

Fig. 4. (a) Measure of the OEO oscillating power as a function of optical power, where three distinct regions of operation are identified. (b) Representative spectra of the OEO frequency in the three regions of operation. Note the asymmetry of the noise sidebands, particularly for Region 1, that is a hallmark of amplitude-phase noise correlation. (c) amplitude noise (AM) and phase noise (PM) in the three operation regions. The laser RIN is shown for comparison. (d) Phase noise of the feedforward-corrected beat note without input optical power optimization compared to operation within Region 2.

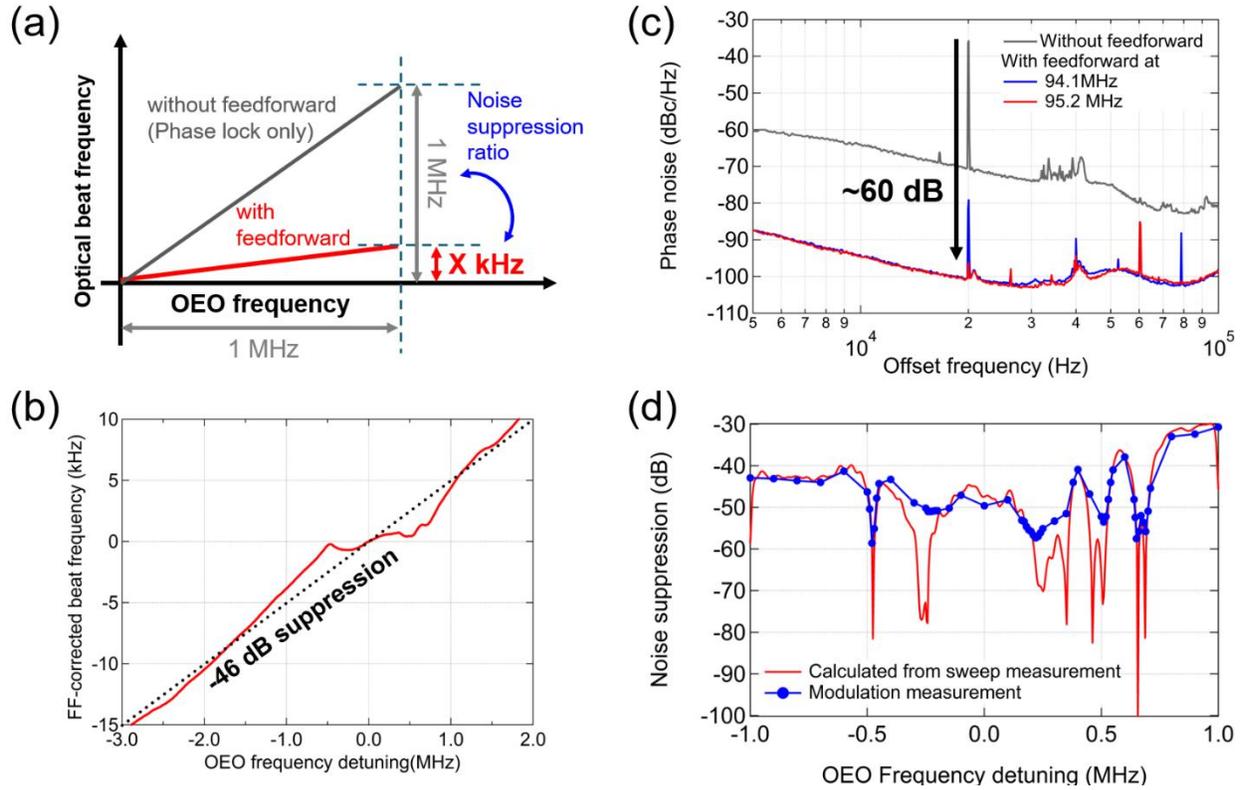

Fig. 5. Direct laser noise suppression ratio measurements. (a) OEO frequency-sweep measurement concept. (b) Measured change in the feedforward-corrected beat frequency as the OEO frequency is tuned. The dotted line indicates the slope that corresponds to 46 dB of suppression of the laser noise. Since the optical power resonant with the FP mode will vary with OEO frequency detuning, the transmitted cavity power was stabilized for this measurement. (c) Measurement of the laser phase noise suppression with RF feedforward using a discrete phase modulation tone. When the OEO oscillation frequency is 95.2 MHz, the tone is suppressed ~60 dB, about 15 dB more than predicted by the linearized noise model. (d) Noise suppression comparison between OEO sweep and discrete tone measurements.

*3.4 DFB laser locking*

The high suppression of laser noise of the OEO laser lock implies that the use of noisier chip-scale laser sources can still achieve cavity noise-limited performance. To demonstrate this, we exchanged the narrow linewidth fiber laser with a compact semiconductor DFB laser in a standard 14-pin butterfly package. The experimental schematic is shown in Fig. 6(a). The DFB laser linewidth is 550 kHz, determined by the beta-separation method from the measured phase noise, shown in the black line in Fig. 6(b). With free-running phase noise ~30 dB (1000×) higher than that of the fiber laser, attempts at direct PDH locking of the DFB laser were unsuccessful. On the other hand, with the OEO locking method, the DFB laser was easily phase locked. For this, a divide-by-four frequency divider was employed, improving the lock robustness and ensuring cycle slip-free operation. Importantly, frequency division for improved robustness is not an option for PDH since the frequency correction signal is directly written onto a baseband voltage. The OEO-locked DFB laser phase noise without feedforward correction is shown in the blue curve of Fig. 6(b). The phase noise at offset frequencies above ~5 kHz is limited by the phase locking feedback bandwidth and gain. Remarkably, the residual laser noise at 10 Hz offset is reduced 140 dB from the free-running laser noise. The residual phase noise at 10 kHz-1 MHz offset was greatly suppressed by RF feedforward, shown in the red curve.

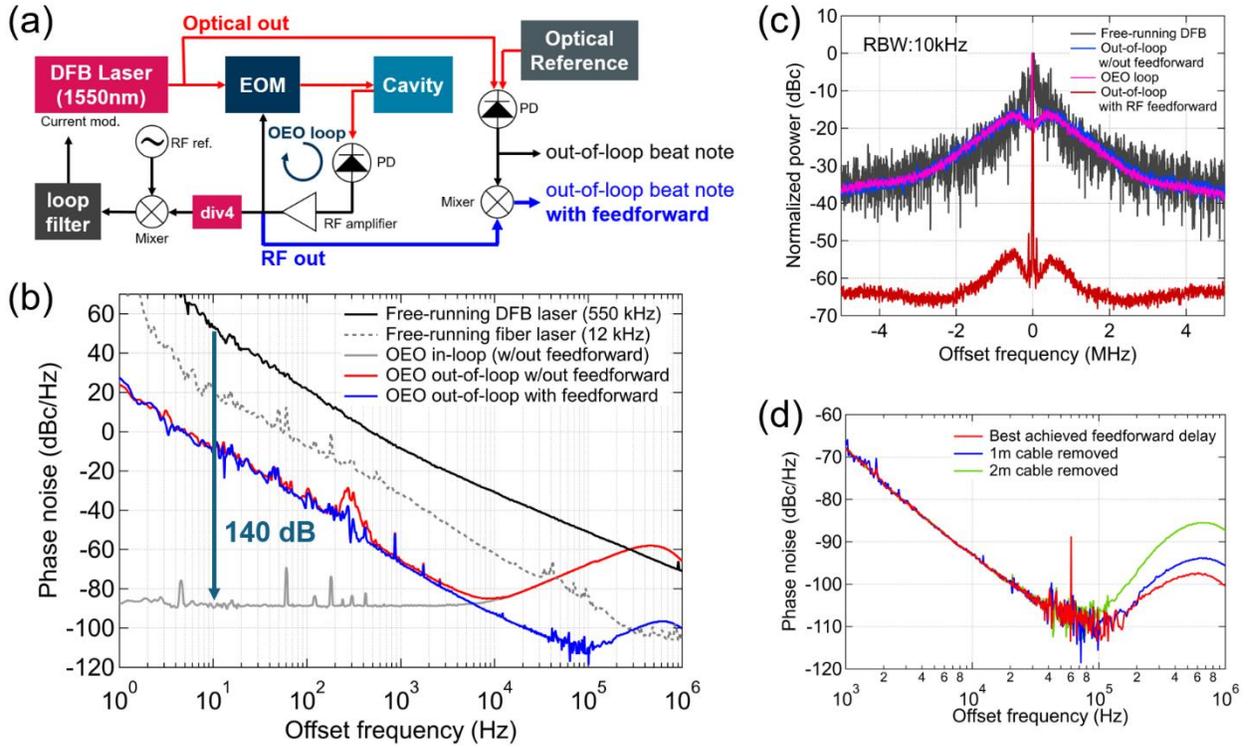

Fig. 6. OEO locking of a DFB laser to an ultrastable FP cavity. (a) Locking setup. (b) Phase noise of the locked DFB laser (blue), phase noise of the locked DFB laser with RF feedforward (red), residual laser noise of the DFB laser when locked (light grey), free-running noise of the DFB (black), and, for comparison, the free-running noise of the fiber laser from Fig. 2(b) (dotted grey). (c) Spectra of the free-running DFB laser (grey), locked DFB without feedforward correction (blue), the OEO oscillation frequency (pink) and the locked DFB with feedforward correction (red). (d) Phase noise of the feedforward-corrected DFB laser as a function of the delay between the heterodyne beat and the applied correction signal.

RF feedforward suppresses the DFB laser noise over a much broader frequency range than is shown in the phase noise curves. Figure 6(c) shows the RF spectra of the direct heterodyne beat of the DFB laser against an ultrastable laser reference over a wider 10 MHz span under various operating conditions, as well as the OEO oscillation frequency. Note that the center frequencies of the various signals have been aligned in post processing to ease comparison of the line shapes. Locking the DFB laser to the cavity results in a coherent peak, though with significant noise sidebands. The noise of the OEO oscillating frequency closely matches the laser noise as expected, such that the noise subtraction with feedforward greatly reduces the noise of the corrected signal over a broad frequency range. The noise at 5 MHz offset is reduced more than 20 dB with RF feedforward, a level of suppression that is impossible with optical feedforward with an AOM, given the AOM's limited response time.

To achieve high feedforward noise suppression at high offset frequencies, the delay between the OEO frequency and the heterodyne beat must be matched correctly. As shown in Fig. 4(d), the amount of the noise reduction was changed by varying the length of the RF cable used for feedforward. The best noise reduction we realized was 35 dB at 1 MHz, corresponding to a delay error of 50 cm. In this measurement, two feedforward paths are necessary for the cross-correlation-based phase noise measurements. However, the length optimization was only implemented on a common feedforward line before their separation for measurement. Therefore, further length optimization for each feedforward path might further improve the phase noise. For example, to achieve 60 dB suppression at 1 MHz offset, only a 3 cm error in the cable length is acceptable. An analysis of the noise suppression limit due to delay mismatch can be found in the Supplement.

## 4. Conclusion

We have demonstrated OEO laser locking using a sub 1 mL, in-vacuum-bonded optical reference cavity with phase noise reaching -120 dBc/Hz at 200 kHz offset, introduced a new RF feedforward technique that attains broadband noise suppression and used it for low noise referencing an optical frequency comb. We have also revealed extremely high levels of laser noise rejection using feedforward that exceed current models by ~15 dB, and we have shown that robust locking of compact DFB lasers to ultrastable cavities is possible without pre-stabilization. Though not explored here, sensitivity to laser RIN, residual amplitude modulation (RAM) in the EOM and spurious etaloning can impact the long-term stability of cavity-stabilized lasers, and greater understanding how these sources of frequency instability manifest in the OEO lock will be quite valuable. Considering the nonlinear behavior of the OEO lock that we have shown here, the impacts of RIN, RAM and spurious etaloning will likely depend on the nonlinear regime of operation. We note, however, that the OEO lock has already been shown to support $10^{-15}$-level fractional frequency instability [29].

Leaving long-term stability aside for now, these results represent a significant advance in the capabilities of OEO laser locking, particularly relevant to out-of-the-lab applications, such as photonic microwave generation and environmental sensing, that require compact and portable ultrastable laser sources. By eliminating the requirement of a laser with low free-running noise, we anticipate further advances of this technique will take advantage of recent developments in the combination of ultrastable reference cavities with photonic integrated circuits [41], resulting in extremely compact and low noise laser sources. In short, the low phase noise and the system simplicity demonstrated in this paper shows great potential for the OEO lock to be the next gold standard for a laser stabilization.

**Funding.**
National Institute of Standards and Technology; Defense Advanced Research Projects Agency, (DARPA GRYPHON program).

**Acknowledgment.**
We thank Andrew Ludlow and the NIST Yb optical clock team for ultrastable reference light, and Jun Ye and JILA silicon cavity team for ultrastable reference light, and Dahyeon Lee, William Groman, and Tanner Grogan for helpful comments on the paper. Product names are given for scientific purposes only and do not represent an endorsement by NIST.

**Disclosures.**
The authors declare no conflicts of interest.

**Data availability.**
Data underlying the results presented in this paper are not publicly available at this time but may be obtained from the authors upon reason- able request.

**Supplemental document.**
See Supplement for supporting content

**Supplemental document for**
**From ultra-noisy to ultra-stable: optimization of the optoelectronic laser lock**

Takuma Nakamura[1,2], Yifan Liu[1,2], Naijun Jin[3], Haotian Cheng[3], Charles McLemore[1], Nazanin Hoghooghi[1], Peter Rakich[3] and Franklyn Quinlan[1,4,5]

[1]Time and Frequency Division, National Institute of Standards and Technology, 325 Broadway, Boulder, CO 80305, USA
[2]Department of Physics, University of Colorado Boulder, 440 UCB Boulder, CO, 80309, USA
[3]Department of Applied Physics, Yale University, New Haven, Connecticut 06520, USA
[4]Electrical, Computer and Energy Engineering, University of Colorado, Boulder, Colorado 80309, USA
[5]fquinlan@nist.gov
*takuma.nakamura@nist.gov

## 1. Theoretical analysis of the OEO laser lock

A steady-state, linearized model that relates the noise of the OEO oscillation frequency to the noise of free-running laser and the noise of the FP-resonant optical sideband has been proposed in [1]. Here we present a slightly different approach intended to complement [1] that reaches the same general conclusion.

A schematic of the model is shown in Fig. S1. We represent the optical field incident on the EOM as

$$E_L(t) = E_0 e^{i\Omega_L t + i\varphi_L} , \qquad (S1)$$

where $E_L$ and $E_0$ are an electric field of the free-running laser and its amplitude, respectively, $\Omega_L$ is the laser angular frequency, and $\varphi_L$ is a time-varying, stochastic phase, representing the laser phase noise. The electrical port of the EOM is driven sinusoidally at an angular frequency of $\Omega_{OEO}$, whose phase is $\varphi_{OEO}$. We will assume small-signal modulation by the EOM, such that the optical field after the EOM consists of a carrier and two sidebands only, given as

$$E_{EOM}(t) = E_0 J_0(\beta) e^{i\Omega_L t + i\varphi_L} + E_0 J_1(\beta) e^{i(\Omega_L + \Omega_{OEO})t + i\varphi_L + i\varphi_{OEO}} - E_0 J_1(\beta) e^{i(\Omega_L - \Omega_{OEO})t + i\varphi_L - i\varphi_{OEO}} . \qquad (S2)$$

For Eq. (S2), $\beta$ is the EOM modulation index, and $J_i(\beta)$ for $i = 0 \ldots 2$ are, respectively, the zeroth, first and second order Bessel functions of the first kind. When $\Omega_{OEO}$ is a resonant frequency of the OEO, $\varphi_{OEO}$ is equal to an integer multiple of $2\pi$ after a full roundtrip around the OEO loop. We therefore propagate the optical and electrical signals around the loop and set the final RF phase equal to the initial phase, modulo $2\pi$.

The modulated optical field is directed to the FP cavity, where we choose the lower sideband at frequency $\Omega_L - \Omega_{OEO}$ to resonate with an FP cavity mode. In this case, the reflected field $E_{refl}$ is

$$E_{refl} = F(\Omega_L)E_0 J_0(\beta)e^{i\Omega_L t+i\varphi_L} + F(\Omega_L + \Omega_{OEO})E_0 J_1(\beta)e^{i(\Omega_L+\Omega_{OEO})t+i\varphi_L+i\varphi_{OEO}} - E_0 J_1(\beta)F(\Omega_L - \Omega_{OEO})e^{i(\Omega_L-\Omega_{OEO})t+i\varphi_L-i\varphi_{OEO}},  \quad (S3)$$

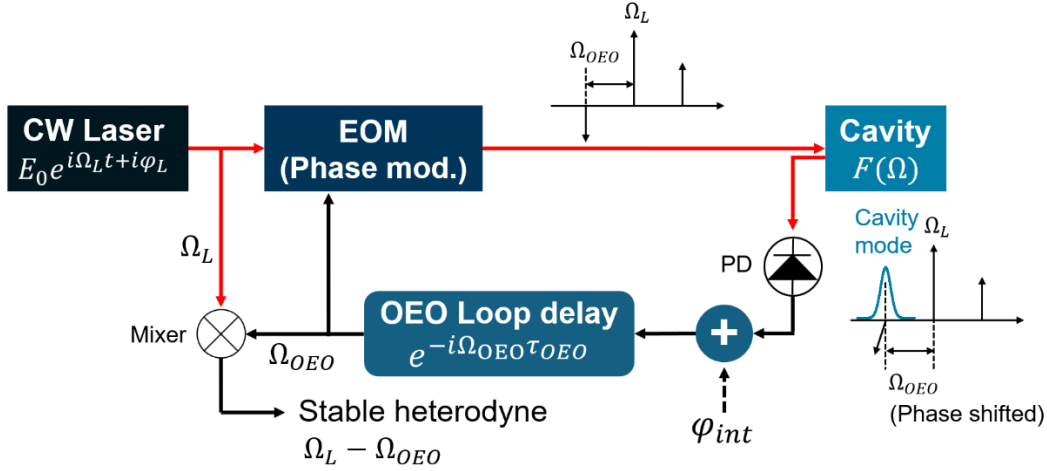

Fig. S1. OEO laser lock noise model. In steady-state operation, the laser frequency is modulated by an electro-optic phase modulator, creating discrete sidebands. The lower frequency sideband is resonant with a mode of the Fabry-Perot (FP) cavity, resulting in a phase shift. The optical field reflected by the cavity is photodetected and used to drive the phase modulator, creating a regenerative opto-electronic oscillator (OEO) loop. Feedforward correction is implemented by mixing the OEO frequency with the laser output. Feedforward correction may be applied optically with an AOM, or in the RF domain after the laser is heterodyned with a separate source.

where $F(\Omega)$ is the cavity response at frequency $\Omega$. Since only the lower sideband is resonant, and the laser carrier and upper sideband frequencies are far detuned, $F(\Omega_L)$ and $F(\Omega_L + \Omega_{RF})$ can be set to unity to very good approximation. To represent the photodetected signal, $i(t)$, we take the square-modulus of Eq. (S3), yielding an analytic signal of

$$i(t) = P_c + P_s(1 + |F(\Omega_L - \Omega_{OEO})|^2) + \sqrt{P_c P_s}(1 - F^*(\Omega_L - \Omega_{OEO}))e^{i\Omega_{OEO}t+i\varphi_{OEO}} + \mathcal{O}(2\Omega_{RF}), \quad (S4)$$

where $F^*(\cdot)$ denotes complex conjugate of $F(\cdot)$, and we have used the following:

$$P_c = |E_0 J_0(\beta)|^2 \quad (S5a)$$

$$P_s = |E_0 J_1(\beta)|^2 \quad (S5b)$$

$$\sqrt{P_c P_s} = |E_0|^2 J_0(\beta)J_1(\beta). \quad (S5c)$$

The photocurrent contains terms at DC, and $\Omega_{RF}$, as well as higher order terms at $2\Omega_{RF}$ that are denoted in Eq. (S4) as $\mathcal{O}(2\Omega_{RF})$. Phase noise from the OEO loop, $\varphi_{int}$, such as photocurrent shot noise or noise from electrical amplification, is then added, followed by the phase associated with the OEO loop propagation delay, $\tau_{OEO}$, given by $\Omega_{OEO}\tau_{OEO}$. The photocurrent term of Eq. (S4) that oscillates at $\Omega_{OEO}$ then becomes

$$i_{RF}(t) = \sqrt{P_c P_s}(1 - F^*(\Omega_L - \Omega_{OEO}))e^{i\Omega_{OEO}t+i\varphi_{OEO}+i\varphi_{int}-i\Omega_{OEO}\tau_{OEO}}. \quad (S6)$$

For $F^*(\Omega_L - \Omega_{OEO})$, we follow [2] and represent the high finesse cavity response as:

$$F^*(\Omega_L - \Omega_{OEO}) = \frac{\xi - i\Omega_o/\Omega_p}{1 - i\Omega_o/\Omega_p}, \quad (S7)$$

where $\Omega_o/2\pi$ is the frequency detuning from the center of the cavity resonance, $\Omega_p/2\pi$ is the half-width at half maximum of the cavity's transmitted power (that is, one-half the cavity linewidth), and $\xi$ is the reflectivity of the electric field on resonance. Perfect spatial mode-matching has been assumed. The value of $\xi$ can range from $-1 \leq \xi \leq 1$; critical coupling is when $\xi = 0$, under-coupled cavities yield $0 < \xi \leq 1$, and over-coupled cavities yield $-1 \leq \xi < 0$. For cavities with symmetric mirror transmission and loss, as is typically the case for ultrastable Fabry-Pérot optical reference cavities, the cavity is always under-coupled.

With Eq. (S7), we obtain

$$1 - F^*(\Omega_L - \Omega_{OEO}) = \frac{1-\xi}{1+\Omega_o^2/\Omega_p^2}\left(1 + i\,\Omega_o/\Omega_p\right). \tag{S8}$$

When the sideband frequency is close to the resonance such that $|\Omega_o| \ll \Omega_p$, the phase of this term is simply $\Omega_o/\Omega_p$. We may then recast Eq. (S6) as

$$i_{RF}(t) = \sqrt{P_c P_s}\,\frac{1-\xi}{1+\Omega_o^2/\Omega_p^2}\sqrt{1+\Omega_o^2/\Omega_p^2}\,e^{i\Omega_{OEO}t + i\varphi_{OEO} + i\varphi_{int} - i\Omega_{OEO}\tau_{OEO} + i\Omega_o/\Omega_p}. \tag{S9}$$

In steady-state operation, the phase after one round-trip around the loop must equal the initial phase, modulo $2\pi$. The phase condition for the OEO frequency at $\Omega_{OEO}$ then becomes

$$\varphi_{OEO} = \varphi_{OEO} + \Omega_o/\Omega_p - \Omega_{OEO}\tau_{OEO} + \varphi_{int}. \tag{S10}$$

The cavity half-linewidth may be written in terms of the cavity lifetime. The cavity lifetime, $\tau_0$, is defined as the ringdown time of the intensity:

$$I(t) = I_0 e^{-t/\tau_0}. \tag{S11}$$

The relationship between the cavity lifetime and the cavity linewidth is given by

$$\tau_0 = \frac{1}{2\Omega_p}. \tag{S12}$$

We can instead define a field ringdown time, $\tau_{cav}$, given by

$$E(t) = E_0 e^{-t/\tau_{cav}}. \tag{S13}$$

Since $I(t) \propto E^2(t)$, it is easy to show

$$\tau_{cav} = 2\tau_0 = \frac{1}{\Omega_p}. \tag{S14}$$

As discussed in [1], $\tau_{cav}$ may also be interpreted as the group delay of the cavity electric field.

The detuning from the FP cavity resonance can be written in terms of the laser frequency, OEO frequency and cavity resonant frequency as

$$\Omega_o = \Omega_L - \Omega_{cav} - \Omega_{OEO}. \tag{S15}$$

Substituting Eq. (S14) and Eq. (S15) into Eq. (S10) yields

$$\Omega_{OEO} = \frac{\tau_{cav}}{\tau_{cav}+\tau_{OEO}}(\Omega_L - \Omega_{cav}) + \frac{\varphi_{int}}{(\tau_{cav}+\tau_{OEO})}. \tag{S16}$$

When the first term of Eq. (S16) dominates, OEO frequency is given by the separation of the laser carrier frequency from the resonant cavity mode, scaled by the ratio of the OEO loop delay to the cavity field ringdown time. The second term of Eq. (S16) represents a phase shift internal to the OEO loop that results in a proportional frequency shift of the OEO oscillation.

To find the power spectral density (PSD) of the frequency fluctuations, $S^y(f)$ (units: Hz²/Hz), we take the Fourier transform of the autocorrelation function of $\Omega_{OEO}/2\pi$ [3]. This yields

$$S^y_{OEO}(f) = \left(\frac{\tau_{cav}}{\tau_{cav}+\tau_{OEO}}\right)^2 \times \left(S^y_L(f) + S^y_{cav}(f)\right) + \frac{S^\phi_{int}(f)}{(2\pi)^2(\tau_{cav}+\tau_{OEO})^2}, \tag{S17}$$

where we have assumed fluctuations on $\Omega_L$, $\Omega_{cav}$ and $\varphi_{OEO}$ are all uncorrelated. This frequency noise can be converted to phase noise by way of $S^\phi(f) = S^\gamma(f)/f^2$, such that the phase noise of the OEO lock is given by

$$S^\phi_{OEO}(f) = \left(\frac{\tau_{cav}}{\tau_{cav}+\tau_{OEO}}\right)^2 \times \left(S^\phi_L(f) + S^\phi_{cav}(f)\right) + \frac{S^\phi_{int}(f)}{(2\pi f)^2(\tau_{cav}+\tau_{OEO})^2} \quad . \tag{S18}$$

For feedforward correction, $\Omega_{OEO}$ is subtracted from the laser frequency:

$$\Omega^{FF}_{OEO} = \Omega_L - \Omega_{OEO} = \frac{\tau_{OEO}}{\tau_{cav}+\tau_{OEO}}\Omega_L - \frac{\tau_{cav}}{\tau_{cav}+\tau_{OEO}}\Omega_{cav} + \frac{\varphi_{OEO}}{(\tau_{cav}+\tau_{OEO})} \quad . \tag{S19}$$

The resulting phase noise PSD of the feedforward-corrected signal is

$$S^{\phi,FF}_{OEO}(f) = \left(\frac{\tau_{OEO}}{\tau_{cav}+\tau_{OEO}}\right)^2 S^\phi_L(f) + \left(\frac{\tau_{cav}}{\tau_{cav}+\tau_{OEO}}\right)^2 S^\phi_{cav}(f) + \frac{S^\phi_{int}(f)}{(2\pi f)^2(\tau_{cav}+\tau_{OEO})^2}. \tag{S20}$$

When $\tau_{cav} \gg \tau_{OEO}$, Eq. (20) reduces to

$$S^{\phi,FF}_{OEO}(f) \approx \left(\frac{\tau_{OEO}}{\tau_{cav}}\right)^2 S^\phi_L(f) + S^\phi_{cav}(f) + \frac{S^\phi_{int}(f)}{(2\pi f)^2(\tau_{cav})^2} \quad . \tag{S21}$$

Thus, with feedforward correction, the cavity and OEO phase noise remain while the laser noise is suppressed. We stress that Eqs. (S18), (S20) and (S21) hold for offset frequencies close to the carrier. Extension to larger offset frequencies with numerical modeling is discussed in [1].

## 2. Investigation of nonlinear noise suppression effect

The assumptions of the model of Section 1 – that only the first-order sidebands are significant and that terms at harmonics of $\Omega_{OEO}$ can be ignored – can be violated when the optical power is high or there is large RF gain in the OEO loop. Such high-power conditions can lead to OEO oscillations at harmonic frequencies, creating additional terms that contribute to the phase and amplitude of the fundamental OEO oscillation frequency. While these higher order oscillations can be electrically filtered out, such filtering can increase $\tau_{OEO}$, reducing the level of feedforward noise suppression. Additionally, amplitude-to-phase conversion in the OEO loop and cavity is more significant at higher powers, producing further modifications to the phase noise beyond what is accounted for in Eqs. (S18) and (S20). Amplitude-to-phase conversion was touched on in the main text. In this section, we discuss the influence of an additional OEO oscillation at the second harmonic frequency.

Figure S2(a) shows the change in the feedforward-corrected heterodyne beat as a function of OEO loop frequency sweep (92-97 MHz), which is the exact same data set as shown in the main text. The same measurements were repeated with various input optical powers, but the overall structure did not change significantly, even for powers close to the oscillation threshold. Therefore, this behavior is not strongly related to the input optical power or the OEO RF power. Importantly, the second harmonic of the OEO oscillation also oscillates at a frequency near 190 MHz. Higher-order harmonics are outside the bandwidth of our photodetector. The existence of a strong second harmonic will create additional optical sidebands offset by $2\Omega_{OEO}$. These sidebands then interfere with sidebands generated by the OEO frequency but that are converted to second-order by the nonlinearity of the EOM, also offset from the carrier at $2\Omega_{OEO}$. One of the second-order sidebands then mixes with the optical sideband (offset by $\Omega_{OEO}$) that is resonant with the FP cavity mode, resulting in an additional phase shift of the OEO resonant frequency. The amount of phase shift will depend on the relative power of the fundamental and second harmonic OEO frequencies, as well as the group delay of the second harmonic around the OEO loop.

To examine this further, we compared the behavior of the feedforward-corrected heterodyne signal when operating at a higher OEO frequency. As the OEO can oscillate at every frequency that meets the OEO resonant condition, we increased the laser detuning from the cavity resonance to match the next OEO harmonic near 139 MHz. The frequency sweep measurement was then repeated, sweeping from 137 MHz to 142 MHz, shown in Fig. S2(b). Both the measurements shown in Figs. S2(a) and S2(b) use the same input optical power. Compared to operating the OEO near 95 MHz, the slope of the feedforward-corrected signal is more linear when operating near 139 MHz. The second harmonic of the 139 MHz oscillation is also present at ~278 MHz, but with lower power as a result of being outside the photodiode bandwidth of 150 MHz. Comparisons of the OEO oscillation powers when operating at 95 MHz and 139 MHz, as well as their respective second harmonics, is shown in Fig. S2(c). The power of the fundamental OEO oscillation does not change when the OEO oscillation frequency is changed from ~95 MHz to ~139 MHz. In contrast, the second harmonic is suppressed by >5 dB when the OEO frequency was around 139 MHz. Note that the data in Fig. S2(c) were taken using a 10 dB RF coupler before the EOM with a RF spectrum analyzer using the max-hold function during periodic sweeping. We examined the power ratios between the fundamental and the second harmonic for both cases, shown in Fig. S2 (d) and (f). Interestingly, as the OEO frequency is swept, the power ratios of the first and second harmonics display a similar structure to that of the feedforward corrected beat signals. While we do not consider this as direct proof of the relationship between the second harmonic and the nonlinear noise suppression, the correlation is compelling, such that further analysis and measurements in the future are warranted.

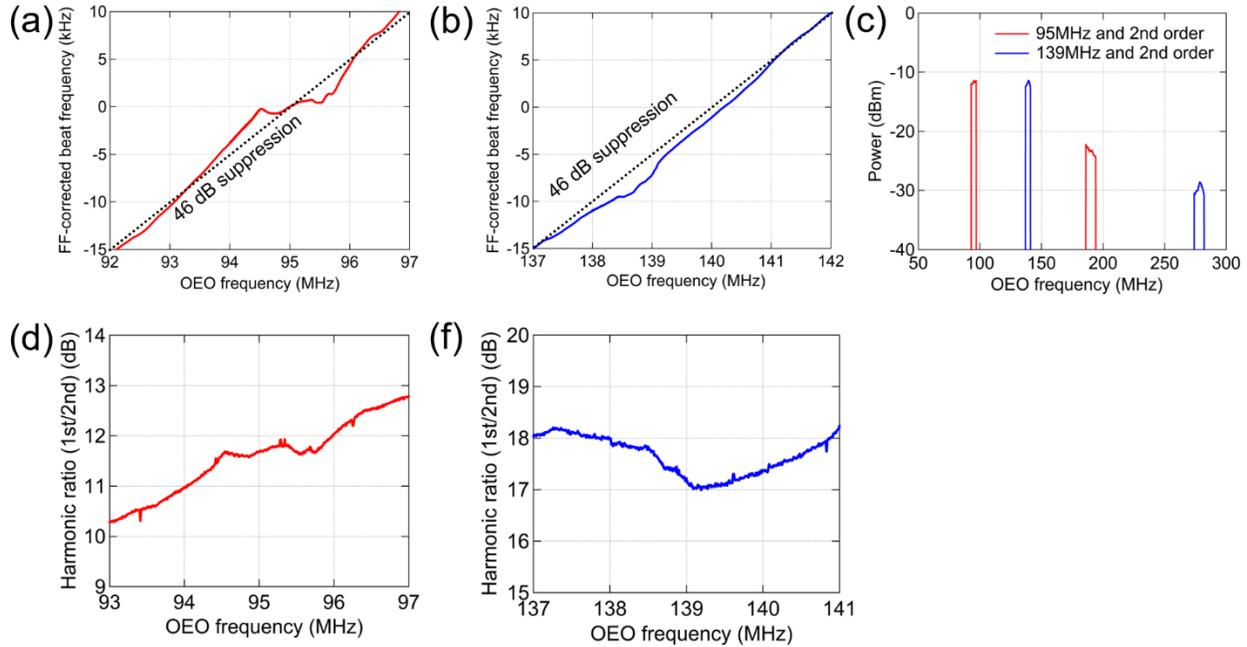

Fig. S2. Nonlinear behavior of the OEO laser lock. Change in the feedforward-corrected beat frequency as the OEO frequency is swept for an OEO oscillation (a) near 95 MHz and (b) near 139MHz. (c) RF powers of the fundamental and second harmonic OEO frequencies for OEO oscillations at ~95 MHz and ~139 MHz. Power ratio of the fundamental and second harmonic OEO frequencies as the OEO frequency is swept (d) near 95 MHz and (f) near 139 MHz.

## 3. Comparison with PDH

### 3.1 Laser noise suppression

In the main text it is stated that the OEO laser lock with feedback to the laser provides greater noise suppression than PDH locking for low offset frequencies. Here we discuss why this is the case.

Figure S3 shows schematics of the feedback control loop models for PDH and the OEO lock [4,5]. For PDH, shown in Fig. S3(a), the feedback loop consists of a frequency comparison between the laser and reference cavity mode, a discriminator that produces a voltage signal that is proportional to the frequency difference, a loop filter for frequency-dependent gain, and a feedback actuator whose frequency at the output is proportional to the voltage on the input (that is, the laser is equivalent to a voltage controlled oscillator). Various noise inputs are included, representing the free-running laser noise, $f_L$, noise from photodetection and following amplification, $v_D$, and noise from the loop filter, $v_{LF}$.

In steady state, the frequency noise at the system output may be derived by requiring self-consistency after one round-trip around the loop. This yields a frequency domain expression of

$$f_o = KGD(f_{cav} - f_o) + KGv_D + Kv_{LF} + f_L ,\qquad(S22)$$

where $f_o$ is the desired laser output frequency, $f_{cav}$ is the FP cavity frequency, $K$ is the laser actuator gain, $G$ is the loop filter gain and $D$ is the frequency discriminator gain.

Solving for $f_o$ gives

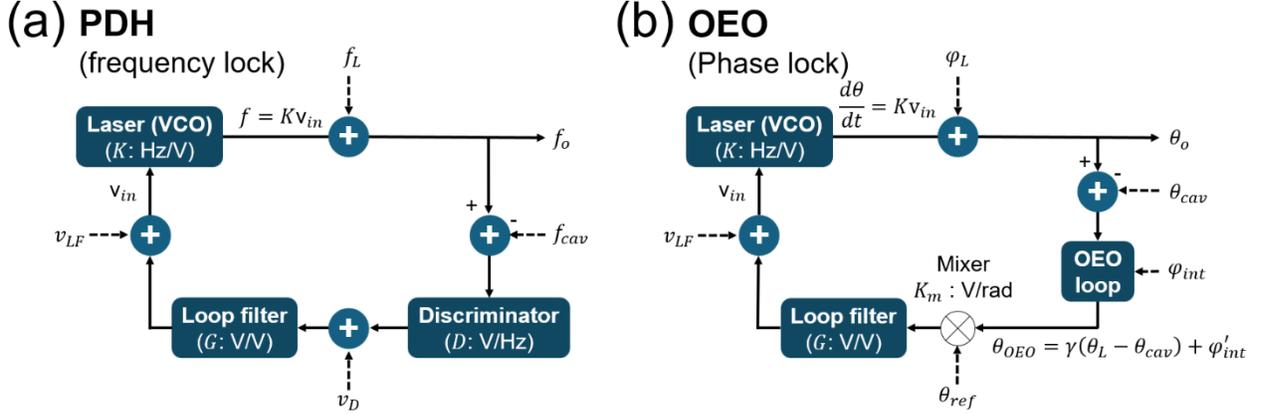

Fig. S3. Models for feedback control. Major system elements and input noise sources are shown, including the laser as a voltage-controlled oscillator (VCO), a loop filter providing frequency dependent gain, and either a frequency discriminator or OEO loop. (a) Pound-Drever-Hall (PDH) laser locking. (b) OEO laser locking.

$$f_o = \frac{KGDf_{cav} + KGv_D + Kv_{LF} + f_L}{1 + KGD} \approx f_{cav} + v_D/D + v_{LF}/GD + f_L/KGD, \qquad(S23)$$

where the approximation is given in the high-gain limit. The output phase noise due to unsuppressed noise of the laser is therefore given by

$$PDH_o^{\phi,L}(f) = \frac{S_L^\phi(f)}{K^2 G^2 D^2}. \qquad(S24)$$

Importantly, the discriminator is frequency dependent [2], and may be written as

$$D = \frac{D_0}{\sqrt{f_p^2 + f^2}}, \qquad(S25)$$

where $f_p = \Omega_p/2\pi$, and $D_0$ is defined as

$$D_0 = 2\sqrt{P_c P_s}(1 - \xi). \qquad(S26)$$

The frequency dependence of $D$ is a consequence of the fact that, within the optical cavity linewidth, $D$ is proportional to the optical frequency, whereas outside the cavity linewidth $D$ is proportional to the optical phase. This transition from frequency to phase proportionality is key to the behavior of the PDH lock.

Using Eq (S23), the expression for $PDH_o^{\phi,L}(f)$ becomes

$$PDH_o^{\phi,L}(f) = \frac{S_L^\phi(f) \times (f_p^2 + f^2)}{K^2 G^2 D_0^2}. \qquad(S27)$$

For offset frequencies well below the optical reference cavity linewidth, the residual noise is

$$PDH_o^{\phi,L}(f) \approx S_L^{\phi}(f) \times \frac{f_p^2}{K^2 G^2 D_0^2}. \tag{S28}$$

Thus, for a given loop filter gain, the close-in laser noise suppression is fixed.

A schematic of the OEO loop is shown in Fig. S3(b). For the OEO feedback loop, mixing the OEO oscillation frequency with a phase reference forms a phase discriminator, and we therefore write the behavior in terms of phase. Similar to PDH, we introduce the free-running phase noise of the laser, $\varphi_L$, the phase noise of the cavity, $\theta_{cav}$, and the noise of the OEO loop (such as photocurrent shot noise), $\varphi_{int}$. As shown in Section 1, the relation of the phase of the OEO frequency, $\theta_{OEO}$, to $\theta_{cav}$, $\varphi_{int}$, and steady-state phase of the laser, $\theta_o$, can be expressed as

$$\theta_{OEO} = \gamma(\theta_o - \theta_{cav}) + \varphi'_{int}, \tag{S29}$$

where we have defined $\gamma$ and $\varphi'_{int}$ as

$$\gamma = \left(\frac{\tau_{cav}}{\tau_{cav}+\tau_{OEO}}\right) \tag{S30a}$$

$$\varphi'_{int} = \frac{\varphi_{int}}{2\pi f(\tau_{cav}+\tau_{OEO})}. \tag{S30b}$$

The OEO additionally has the phase of the reference signal, $\theta_{ref}$. Also, note that the laser feedback actuates on the derivative of the phase. Self-consistency around the OEO phase-lock loop yields a frequency domain expression of

$$KK_m G(\theta_{ref} - \gamma\theta_o + \gamma\theta_{cav} - \varphi'_{int}) + Kv_{LF} = i2\pi f(\theta_o - \varphi_L), \tag{S31}$$

where $K$ is the laser actuator gain, $G$ is the loop filter gain, and $K_m$ is the effective gain of mixing with the phase reference $\theta_{ref}$. In Eq. (S31), we have used the fact that the Fourier transform of a derivative results in multiplication by $i2\pi f$ in the frequency domain. Solving for the laser phase at the output in the high-gain limit yields

$$\theta_o = \theta_{ref}/\gamma + \theta_{cav} - \varphi'_{int}/\gamma + \frac{v_{LF}}{\gamma K_m G} + \frac{\varphi_L}{\gamma KK_m G/(i2\pi f)}. \tag{S32}$$

Phase noise resulting from unsuppressed laser noise becomes

$$OEO_o^{\phi,L}(f) = \frac{S_L^{\phi}(f) \times (2\pi f)^2}{\gamma^2 K^2 K_m^2 G^2}. \tag{S33}$$

Unlike the PDH lock, for a given loop filter gain the residual laser noise continues to see greater suppression as the offset frequency is decreased.

If we assume availability of the same laser, feedback actuator (such as laser cavity PZT or external AOM) and loop filter, the ratio of the unsuppressed laser noise for OEO and PDH becomes

$$\frac{OEO_o^{\phi,L}}{PDH_o^{\phi,L}} = \frac{(2\pi f)^2}{(f_p^2+f^2)} \times \frac{D_0^2}{\gamma^2 K_m^2}. \tag{S34}$$

We note that the mixer constant $K_m$ is proportional to the amplitude of the OEO signal, such that

$$K_m^2 = \alpha \times P_c P_s (1-\xi)^2, \tag{S35}$$

where $\alpha$ is a proportionality constant depending on the amplification of the OEO signal and the mixer loss. The OEO-PDH noise ratio may then be expressed as

$$\frac{OEO_o^{\phi,L}}{PDH_o^{\phi,L}} = \frac{f^2}{(f_p^2+f^2)} \times \frac{(4\pi)^2}{\gamma^2 \alpha} \approx \frac{f^2}{(f_p^2+f^2)} \times \frac{(4\pi)^2}{\alpha}, \tag{S36}$$

where we have assumed $\tau_{cav} \gg \tau_{OEO}$, such that $\gamma \approx 1$. For offset frequencies well below $f_p$, this becomes

$$\frac{OEO_o^{\phi,L}}{PDH_o^{\phi,L}} = \frac{f^2}{f_p^2} \times \frac{(4\pi)^2}{\alpha}. \tag{S37}$$

For offset frequencies well above $f_p$, the ratio becomes fixed:

$$\frac{OEO_o^{\phi,L}}{PDH_o^{\phi,L}} = \frac{(4\pi)^2}{\alpha}. \tag{S38}$$

For the results presented in Fig. 2(b) of the main text, the OEO lock and PDH residual laser noise far from carrier are nearly equal. Equation (S37) then predicts lower residual laser noise for the OEO lock for $f < f_p$. Comparison of the light blue and light red curves of Fig. 2(b) show that this is indeed the case.

In Section 1, we examined the phase noise output of the OEO after feedforward correction. Here we reexamine the feedforward-corrected phase noise with the feedback lock engaged. For this, we solve for $\theta_{OEO}$ in steady state. The system equation in the frequency domain becomes

$$KK_m G(\theta_{ref} - \theta_{OEO}) + K v_{LF} = i 2\pi f(\theta_o - \varphi_L). \tag{S39}$$

The steady-state laser phase can be expressed in terms of $\theta_{OEO}$ as

$$\theta_o = \theta_{OEO}/\gamma + \theta_{cav} + \varphi'_{int}/\gamma. \tag{S40}$$

Substituting Eq. (S40) into Eq. (S39) and assuming the high-gain limit yields

$$\theta_{OEO} = \theta_{ref} + \frac{v_{LF}}{K_m G} - \frac{\theta_{cav}}{KK_m G/(i2\pi f)} - \frac{\varphi'_{int}}{\gamma KK_m G/(i2\pi f)} + \frac{\varphi_L}{KK_m G/(i2\pi f)}. \tag{S41}$$

The feedforward-corrected signal is then

$$\theta_o - \theta_{OEO} = \theta_{ref}\left(\frac{1-\gamma}{\gamma}\right) + \frac{\varphi_L}{KK_m G/(i2\pi f)}\left(\frac{1-\gamma}{\gamma}\right) + v_{LF}/KG\left(\frac{1-\gamma}{\gamma}\right) +$$
$$(\theta_{cav} - \varphi'_{int}/\gamma)\left(1 + \frac{1}{KK_m G/(i2\pi f)}\right). \tag{S42}$$

Using Eq. (40) and assuming high gain, Eq. (S42) simplifies to

$$\theta_o - \theta_{OEO} \approx \theta_{ref}\left(\frac{\tau_{OEO}}{\tau_{cav}}\right) + \frac{\varphi_L}{KK_m G/(i2\pi f)}\left(\frac{\tau_{OEO}}{\tau_{cav}}\right) + \frac{v_{LF}}{KG}\left(\frac{\tau_{OEO}}{\tau_{cav}}\right) + \theta_{cav} - \varphi'_{int}. \tag{S43}$$

The PSD of the feedforward phase noise is then

$$S_{OEO}^{\phi,FF}(f) \approx \left(\frac{\tau_{OEO}}{\tau_{cav}}\right)^2 \frac{S_L^\phi(f)}{K^2 K_m^2 G^2/(2\pi f)^2} + S_{cav}^\phi(f) + \frac{S_{int}^\phi(f)}{(2\pi f)^2 (\tau_{cav})^2} + \left(\frac{\tau_{OEO}}{\tau_{cav}}\right)^2 S_{ref}^\phi(f) + \left(\frac{\tau_{OEO}}{\tau_{cav}}\right)^2 \frac{S_{LF}^v(f)}{K^2 G^2}, \tag{S44}$$

where $S_{LF}^v(f)$ is the voltage noise added by the loop filter. Comparing to Eq. (S21) from Section 1, we note that this expression provides the same level of cavity and intrinsic noise, that the free-running laser noise is reduced by the gain of the feedback loop as expected, and that the phase noise contributions from the phase reference and the loop filter are suppressed by the ratio of $\tau_{OEO}$ to $\tau_{cav}$.

## 3.2 Comparison of PDH and OEO in-loop noise

It is interesting to compare $S_{int}^\phi(f)$ to the in-loop noise of PDH. As in previous sections, we present analytical expressions that complement the transfer function approach and numerical modeling given in [1].

Both the OEO lock and PDH suffer from noise of the detectors and amplifiers, used to generate locking signals for feedback control, that is imparted on the output signal. For PDH, the phase noise from this in-loop noise is determined by the ratio of the noise, denoted $N_{PDH}$ (units: V²/Hz), to the square of the discriminator slope, $D$ (units: Hz/V). From Eqs. (S25) and (S26), the discriminator slope is a function of offset frequency and is given by

$$D(f) = \frac{4\sqrt{P_c P_s}(1-\xi)}{(2f_p)\sqrt{1+f^2/f_p^2}}. \tag{S45}$$

The in-loop noise floor frequency noise PSD for PDH is then

$$PDH_{int}^\gamma(f) = \frac{4 f_p^2 \left(1+\frac{f^2}{f_p^2}\right)}{16 P_c P_s (1-\xi)^2} N_{PDH}. \tag{S46}$$

This simplifies to

$$PDH^{\gamma}_{int}(f) = \frac{N_{PDH}}{4P_cP_s(1-\xi)^2}\left(\frac{1}{(2\pi\tau_{cav})^2} + f^2\right). \quad (S47)$$

The corresponding phase noise is then

$$PDH^{\phi}_{int}(f) = \frac{N_{PDH}}{4P_cP_s(1-\xi)^2}\left(\frac{1}{(2\pi\tau_{cav}f)^2} + 1\right). \quad (S48)$$

For the OEO lock, the phase noise $S^{\phi}_{int}(f)$ represents the in-loop noise, and its noise contribution for offset frequencies less than the cavity half-linewidth is given in Eq. (S25). This expression is extended to larger offset frequencies with a Leeson model description of the OEO [1,6], resulting in

$$S^{\phi}_{OEO}(f) = S^{\phi}_{int}(f)\left(\frac{1}{(2\pi f)^2(\tau_{cav}+\tau_{OEO})^2} + 1\right). \quad (S49)$$

We assume noise from photodetection and RF amplification, $N_{OEO}$, is distributed equally between the amplitude and phase of $\Omega_{OEO}$. In this case, the phase noise $S^{\phi}_{int}(f)$ is given by $N_{OEO}/(2P_{OEO})$ for OEO signal power of $P_{OEO}$.

From Eq. (S9), when the sideband is on resonance of the optical cavity ($\Omega_o = 0$), the power in the OEO electrical signal is

$$P_{OEO} = 2P_cP_s(1-\xi)^2. \quad (S50)$$

The phase noise of the OEO lock is then

$$S^{\phi}_{int}(f) = \frac{1}{2}\frac{N_{OEO}}{2P_cP_s(1-\xi)^2}. \quad (S51)$$

From Eq. (S49), the OEO phase noise due to $S^{\phi}_{int}(f)$ is then

$$S^{\phi}_{OEO}(f) = \frac{N_{OEO}}{4P_cP_s(1-\xi)^2}\left(\frac{1}{(2\pi f)^2(\tau_{cav}+\tau_{OEO})^2} + 1\right). \quad (S52)$$

For $\tau_{OEO} + \tau_{cav} \approx \tau_{cav}$, this expression becomes

$$S^{\phi}_{OEO}(f) \approx \frac{N_{OEO}}{4P_cP_s(1-\xi)^2}\left(\frac{1}{(2\pi\tau_{cav}f)^2} + 1\right). \quad (S53)$$

Thus, for a given system noise, optical power, and EOM depth of modulation, PDH and the OEO lock are expected to yield the same in-loop noise floor. It is important to note, however, that the presence of higher harmonics of $\Omega_{OEO}$ are indicative of a cyclostationary noise process [7,8]. This may lead to an imbalance of the noise distribution between amplitude and phase of $\Omega_{OEO}$, with the majority of the noise contributing to the amplitude quadrature. Furthermore, the OEO lock may not be operated with the same EOM depth of modulation, leading to a different value of the product $P_cP_s$. For PDH, lowest noise operation is achieved with $P_cP_s$ maximized. In contrast, lowest noise operation for the OEO lock may be in the nonlinear regime where laser noise suppression is greatest, leading to a different value for $P_cP_s$.

To understand the same in-loop floor between PDH and OEO intuitively, a simplified setup of the OEO method is helpful, as shown in Fig. S4. If the phase lock is perfect, the signal from the PD to the EOM, as indicated with blue line in the figure, becomes indistinguishable if the signal is from the photodetector in an OEO regenerative loop (as shown on the left of Fig. S4) or an RF reference for PDH (as shown in right in Fig. S4). In practice, the phase lock is almost perfect at low offset frequencies. Therefore, the achievable noise floor at lower frequency between PDH and OEO should not be different, except the nonlinear behavior of the OEO as mentioned above.

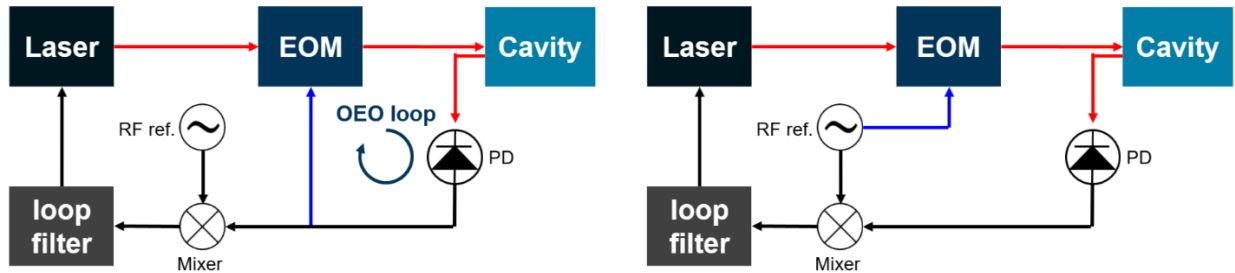

Fig. S4. Simplified schematics of the OEO laser lock (left) and PDH lock (right), stressing the equivalence of the EOM control signals in the high gain limit.

## 4 Delay effect of the feedforward

To achieve the highest level of feedforward noise suppression, the relative delay between the laser signal and the OEO signal must be set appropriately. For instance, 0.01 rad phase error restricts the amount of the noise suppression to around 40 dB. In the RF feedforward case, all optical and RF path lengths from the optical beam splitter to the mixer need to be considered. Figures S5(a) and S5(b) show the delay effects for different offset frequencies and different RF cable length errors. To achieve 60 dB suppression at 1 MHz, the maximum cable length error is only 3 cm. As a result, although a shorter OEO loop (20 cm) used with longer cavity (10 cm) with high finesse (500k) can in principle provide noise suppression of 100 dB=(20log(100μs/1ns)), it is not easy to realize because only 300 μm error is acceptable to achieve 100 dB suppression at 1 MHz.

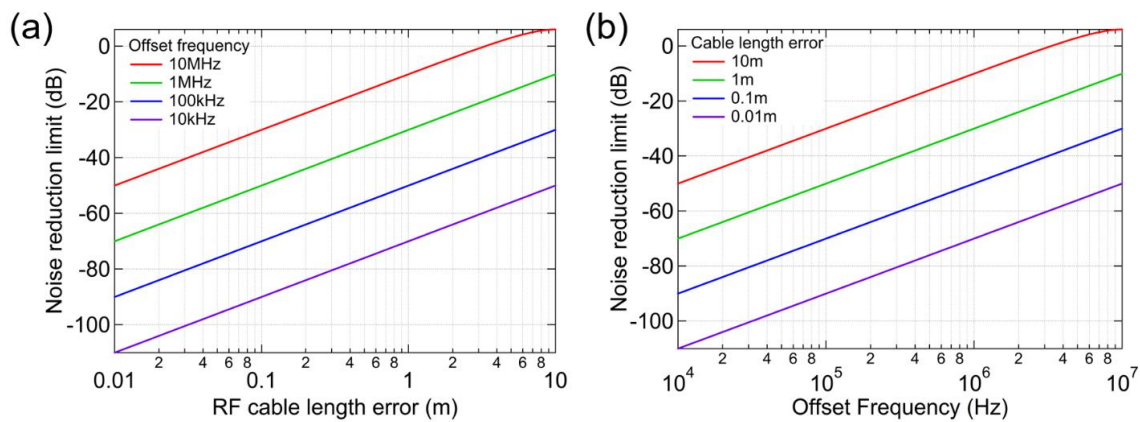

Fig. S5. (a) Limit to the level of feedforward laser noise suppression as a function of RF cable length error for various offset frequencies. (b) Limit to the level of feedforward laser noise suppression as a function of offset frequency for various RF cable length errors.